\documentclass{PoS}

\title{Prospects for neutrino oscillation parameters}

\ShortTitle{Prospects}

\author{\speaker{Patrick Huber}\\
        Center for Neutrino Physics, Virginia Tech, Blacksburg, USA\\
        E-mail: \email{pahuber@vt.edu}}

\abstract{In this contribution we discuss the future of the global
  long-baseline neutrino oscillation program. The case is made that
  our current lack of understanding of neutrino-nucleus interactions is
  a serious challenge which will need to be met with new experimental
  initiatives in neutrino scattering.}

\FullConference{Neutrino Oscillation Workshop\\
		4 - 11 September, 2016\\
		Otranto (Lecce, Italy)}

\begin{document}


\section{Introduction}

The discovery of neutrino oscillation, awarded the 2015 Nobel prize in
physics, is one of the great discoveries of our time. Apart from
finding that neutrinos have mass, we also have been blessed with large
mixing angles in both solar and atmospheric neutrinos as well as a
large value of $\theta_{13}$. All of which are \emph{necessary}
ingredients to allow for the search of CP violation in neutrino
oscillations, since the absolute size of CP effects is suppressed by
the smallest of the mixing angles and by the ratio of two mass squared
differences. The study of CP violation in turn requires the ability to
perform an appearance measurement, and given our lack of proper
technology to efficiently create or detect $\nu_\tau$, this implies
the use of $\nu_e$ and $\nu_\mu$. The need to involve $\nu_\mu$ in
turn requires to use energies in the 100's of MeV and above and this
in turn results in long baselines of 100's km and more.

The event rate for $\nu_e\rightarrow\nu_\mu$ or
$\nu_\mu\rightarrow\nu_e$ appearance in leading order is proportional
to $\sin^22\theta_{13}$ and thus, the fact that $\theta_{13}$ is large
is good news: it allows to obtain a sufficiently large event sample
using conventional pion-decay, horn-focused neutrino beams, which we
have been using for more than 3 decades. It still requires, however,
to push this technology to its limits by using MW-level proton beams
and very large detectors of at least the size of
Super-Kamiokande. These long-baseline experiment are characterized in
terms of megawatts, kilotons and decades. This is a new scale of
effort in neutrino physics and makes neutrino physics, in its sheer
experimental scope and scale, much more similar to traditional
accelerator based science, like for instance the LHC program.

Indeed, DUNE is expected to start data taking roughly a decade from
now and to run for a decade and to absorb a very large fraction of the
resources of its host country and the international neutrino
community. Also for many scientists this will be the only major
experiment during their career. The sheer scale of these new neutrino
experiments also implies that failure to make a major discovery is not
an option. For small-scale neutrino experiments like the ones in the
Booster neutrino beam at Fermilab or reactor neutrino experiments not
finding anything or being preempted by some other measurement is a
completely acceptable outcome because there are many experiments at
this scale. As long as some of these experiments make discoveries
or provide measurements, science and importantly, along with it, the
careers of those conducting these experiments can progress. The
bargain each scientists strikes with a generational experiment like
the LHC is based on the conviction that the science goal is worthy of
a lifetime of struggle and the realization that this is the
\emph{only} way to achieve this particular science goal. Implied in
this bargain is the understanding that both, the worthiness of the
science goal and the unique ability to achieve it by this one specific
experiment, will endure over the decades it takes to carry out this
program.

For the LHC the science goal was the discovery of the Higgs boson and
to find New Physics, either in the form of new degrees of freedom or
by some other breakdown of the Standard Model. It also is understood
that this would require particle collisions at unprecedented energies
and after the global community had settled on the LHC is also was
clear that this will be the \emph{only} machine in this energy regime
for the foreseeable future. The Higgs has been discovered and the
search for New Physics is ongoing, with a major upgrade of the machine
underway -- a roundabout success.

For long-baseline oscillation experiments the science goal is the
discovery of leptonic CP violation and the search for New Physics by
precisely testing the three-flavor oscillation framework. However,
this is where the similarity with the LHC program ends: there is no
international consensus to pursue only one experiment and already
on-going experiments, notably NOvA in the U.S. and T2K in Japan, will
be making inroads into the very question of leptonic CP
violation. While it seems very unlikely that the on-going experiments
will achieve a 5\,$\sigma$ discovery, a 3\,$\sigma$ evidence for
leptonic CP violation may be conceivably obtained. Thus future
long-baseline experiments will not be entering a {\it terra incognita}
but will be tasked with the exploration and mapping of terrain seen
before. Therefore, precision measurements and a comprehensive set of
physics goals is the real scientific objective.

\section{Future prospects}

First tentative hints for leptonic CP violation became apparent
earlier this year~\cite{Marrone} and since more global fits have
been performed~\cite{Esteban:2016qun} indicating a preference for a
value of the leptonic CP phase around $\delta\simeq-\pi/2$. This hint for CP
violation currently is at 1--2$\,\sigma$ level and thus may be nothing
more than a statistical fluctuation. On the other hand, both T2K and
NOvA do consistently report $\nu_e$ appearance rates at the upper end
of the possible range, which is what is expected if $\delta=-\pi/2$.

The T2K collaboration has formulated a plan, including an approved
upgrade of the proton accelerator to increase beam power to 1.2\,MW,
which on a timescale of about a decade will allow T2K to reach a
3\,$\sigma$ rejection of leptonic CP conservation, assuming the
current best-fit value is close to the true value of
$\delta$~\cite{Abe:2016tii}. In combination with continued NOvA running,
current NOvA data represent only about 1/6 of the approved number of
protons on target, and the precise determination of $\theta_{13}$ by
Daya Bay~\cite{An:2016ses} there will a be a good determination of the
leptonic CP phase by 2025. Also, the question whether $\theta_{23}$ is
maximal can be effectively addressed by this data in particular when
combined with atmospheric neutrino data, which Super-Kamiokande
continues to accumulate. Another crucial piece of information is the
neutrino mass hierarchy and while current data seems to have no
particular preference, future data from NOvA has the potential to
answer this question without any ambiguity. Then, of course, there is
JUNO and possibly PINGU all trying to address the same question. It
appears, therefore, likely that the question of the mass hierarchy
will be settled before too long.

\begin{figure}
\begin{center}
\includegraphics[width=0.75\textwidth]{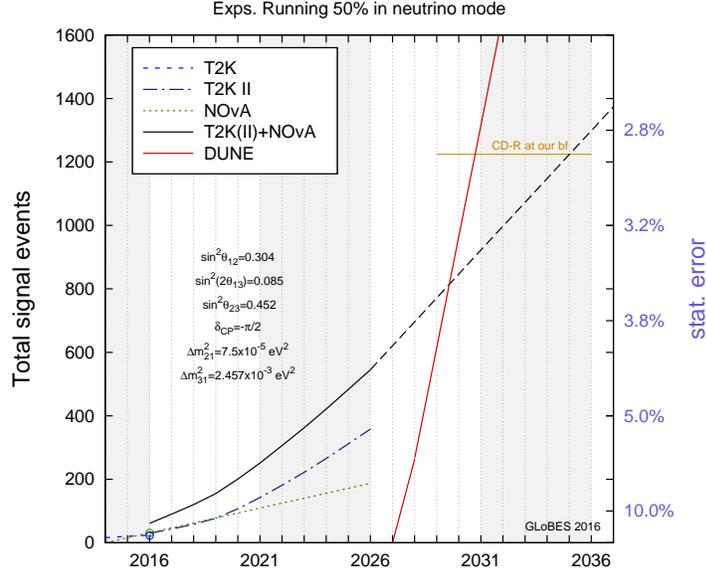}
\end{center}
\caption{\label{fig:time} Shown is the time evolution of the total
  $\nu_\mu\rightarrow\nu_e$ appearance rate for the experiments named
  in the legend. The assumption is that $\delta=-\pi/2$ and the mass
  hierarchy is normal. The axis label on the right-hand vertical axis
  corresponds to $1/sqrt{N}$, with $N$ being the number of events
  shown on the left-hand vertical axis and thus, represents the
  statistical accuracy of the data set. The DUNE run plan is based
  on~\cite{Acciarri:2015uup,Alion:2016uaj,DUNE-meeting:2016}. The NOvA
  run plan is based on \cite{nova-neutrino:2016}. The T2K run plan is
  based on~\cite{Abe:2016tii}.}
\end{figure}
A reasonably good proxy for the actual physics reach of a
long-baseline experiment is given by the total number of events it
accumulates in the $\nu_\mu\rightarrow\nu_e$ appearance channel. 
The time evolution is shown in Fig.~\ref{fig:time}, assuming current
best fit values: the next decade will see an order of magnitude
increase in these numbers and the statistical errors will drop below
5\%. DUNE has to run for about 3 years to double the global event
sample in the appearance channel and even at the end of its planned
run it only will have roughly tripled the available global data.
Also, by the time DUNE data starts to dominate the global data set,
this global data set will comprise about 1500 events, with a
corresponding statistical accuracy of 2.6\%. This excellent
statistical accuracy has to be matched by a corresponding and ideally
somewhat smaller systematic uncertainty -- otherwise the massive
investment in these experiments is wasted.

\section{Road to precision}

There are two challenges for a precision long-baseline experiment:
determining the appearance rate and determining the neutrino
energy. The rate uncertainty is driven by the knowledge of the beam
flux and neutrino interaction cross section. Using a pion-decay,
horn-focused beam, the flux knowledge is at best at the 10\% level
despite many years of efforts to improve the state of the art by
hadron production measurements. In realization of this limitation all
long-baseline experiments employ a near detector or a suite of near
detectors. Following the argument of Ref.~\cite{Huber:2007em}, the
problem lies in the fact that the near detector, at best, measures the
product of beam flux and cross section and thus to know either one, the
other quantity is required; there are fewer observables than
unknowns. In reactor experiments like Daya Bay the problem is solved
by measuring the product of the \emph{identical} cross section, in
this case for inverse beta decay, and the beam flux, in this case
$\bar\nu_e$, in the near and far detectors, which have the same
physical dimensions and characteristics within a fraction of a
percent. In inverse beta decay a unique and clean flavor tag is
obtained and there is no doubt about the underlying micro-physics,
since neutron decay is very well understood.  The source is a point
source and thus the geometrical acceptance difference between near and
far detectors is given by the square of the ratio of baselines, which
in turn can be measured with centimeter precision. Thus, the
oscillation probability can be extracted by simply taking the ratio of
far to near detector data corrected for geometric acceptance.

In a long-baseline experiments \emph{none} of the conditions which
allowed Daya Bay to succeed is met: cross sections in near and far
detectors are different since the neutrino flavor and energy
distribution is different. The beam flux and flavor composition seen
by the near detector is different even in absence of oscillation
because of the complicated acceptance. The near and far detector are
not the same size, do not use the same technology and sometimes not
even the same target material. The near detector sees the decay pipe
as line source whereas the far detector sees it as point
source. Relating the observable quantities to the underlying
micro-physics in an event requires a precise understanding of the
micro-physics, which we lack. Instead, we rely on Monte Carlo
simulations, which have been tuned to existing data, for event
identification. The limitations of this approach are exemplified by
the fact that basically no two cross section measurements performed in
a neutrino beam ever seem to to agree with each other. 

Finally, the need to reconstruct the neutrino energy precisely is a
design feature of experiments using a wide-band neutrino beam and all
the physics benefits of covering a range of $L/E$-values rely on this
reconstruction. Again, the lack of an understanding of the
micro-physics of neutrino-nucleus interactions prevents accurate
neutrino energy reconstruction because the observable signatures do not
have an understood relation to neutrino energy. It has been shown that
approximate schemes like the exploitation of quasi-elastic kinematics
do not provide sufficient accuracy for the next generation of
experiments~\cite{Coloma:2013tba,Ankowski:2015jya} and also
calorimetric methods have their limitations~\cite{Ankowski:2015kya},
in particular with respect to neutral secondary particles like
neutrons. In particular, for the determination of CP violation this is
a major issue~\cite{Coloma:2013rqa}.

Therefore, even an ideal near detector seems to be insufficient to
resolve the systematics problem in long-baseline experiments since it
ultimately provides fewer observables than unknowns. The hope that the
multitude of different event types, charged current quasi elastics,
charged current single pion etc. will provide a sufficient number of
observables is naive: they only can constrain each other if we connect
them with a micro-physical model, which we do not have. It is a fallacy
to confuse existing Monte Carlo event generators with an actual
understanding of the micro-physics as borne out by the great difficulty
to reconcile any new measurement of exclusive neutrino cross sections
with existing ones, to quote from the most recent MINERvA
publication~\cite{DeVan:2016rkm}: ``\emph{Unlike} the
  measurements of the individual processes (quasi- elastic, pion
  production) the total cross section measurements agree with the
  GENIE simulation and prior data to within their uncertainties
  [\ldots]'' (emphasis added). Given that the MINERvA experiments in
many ways represents the state of the art in neutrino scattering,
this is discouraging, despite the success for inclusive cross
sections: neutrino energy reconstruction desperately relies on an
understanding of the exclusive interaction channels.

Usually when the situation seems hopeless on the experimental side, we
turn to theory to provide the needed answers. It is obvious that
describing bound state multi-nucleon systems in the ground state is a
daunting task and so far can be only achieved by using
phenomenological Hamiltonians and for nuclei lighter than A=12. The
problem at hand is however not to know the ground state (or the
low-lying excitations) but the response to energy transfers
well into the GeV-range, which requires a relativistic treatment.
Once the hard scattering event has taken place we also need to
understand how the reaction products get out of the nucleus, a problem
typically summarized under the term final-state
interactions. Recently, Benhar~\cite{Benhar:2016cmq} points out that
many different calculational approaches based on very different
assumption seem to yield the same result, which is puzzling. The role
of electron scattering data can not be overstated, since we can
exploit fully exclusive kinematics to separate the various
micro-physical contributions and \emph{any} model of neutrino
interactions \emph{must} reproduce electron scattering
data. Fortunately, a program is underway to obtain this crucial data
for argon~\cite{Benhar:2014nca}.

\section{Summary}

We are very fortunate in neutrino physics having found neutrino
oscillation, and by association that neutrinos have a mass; also, we
find large mixing angles, including a quite sizable value of
$\theta_{13}$. Thus, the stage is set to study genuine three-flavor
effects and to, hopefully, find leptonic CP violation. There is a
vibrant ongoing experimental effort, spearheaded by NOvA and T2K and
we have won approval for DUNE. The future of neutrino physics lies in
\emph{precision} studies of long-baseline neutrino oscillation and
while a large $\theta_{13}$ allows to accumulate significant event
samples we also will need concomitant control of systematical
uncertainties.

The choice of pion-decay, horn-focused neutrino beams implies poor
knowledge of the primary neutrino flux, which combined with our lack
of understanding of neutrino-nucleus interactions presents a
challenge. In this note we reiterate the argument previously made in
the literature that even a capable near detector complex can not meet
this challenge. The state of theory is such that it is at best unclear
whether theory can provide the missing answers. Let us assume, that
there will a be a breakthrough in theory providing a full model of
neutrino-nucleus interaction including final state interactions. This
model will not be based on the Standard Model Lagrangian or any other
first-principles calculations; given the complexity of the problem it
has to be based on phenomenological insights and appropriate
approximations. Thus, even if we all agree that we have the ``right''
model, we will need to test this model at the level of accuracy we
intend to use it at. Therefore, the neutrino community needs to seriously think
about an experimental neutrino scattering program to accompany the
long-baseline oscillation program, see for
instance~\cite{Adey:2013pio,Bhadra:2014oma}.


\section*{Acknowledgments} This work was supported by the U.S. Department of
Energy Office of Science under award number \protect{DE-SC0009973}.

\bibliographystyle{JHEP} 
\bibliography{references}
\end{document}